\documentclass[]{scrartcl}
\usepackage{PackageManager}

\usepackage[T1]{fontenc}

\KOMAoption{abstract}{true}
\KOMAoption{twocolumn}{off}
\KOMAoption{parskip}{half-}
\recalctypearea

\usepackage{hyperref}
\hypersetup{
	colorlinks  = false, 
	linktocpage = true, 
	breaklinks  = true, 
}
\usepackage[
	round, 
	sort, 
	comma, 
	authoryear, 
	]{natbib}
\title{
A lightweight numerical model for predictive control of \\ borehole thermal energy storages
}
\date{July 2025}

\author{
    Johannes van Randenborgh, Steffen Daniel, and Moritz Schulze Darup
    \thanks{
        J. van Randenborgh, S. Daniel, and M. Schulze Darup are with the Control and Cyberphysical Systems Group, Faculty of Mechanical Engineering, TU Dortmund University, Germany
        {\ttfamily\small \{<first name>.<last name>\}@tu-dortmund.de}.
        This manuscript is accepted for publication at the 5th Modeling, Estimation and Control Conference (MECC 2025) in Pittsburgh, Pennsylvania, USA.
    }
}

\titlehead{
        $\copyright$ to the authors.
        This work has been accepted to IFAC for publication under a Creative Commons Licence CC-BY-NC-ND.
}
\begin{document}

\maketitle

\begin{abstract}
Borehole thermal energy storage (BTES) can reduce the operation of fossil fuel-based heating, ventilation, and air conditioning systems for buildings.
With BTES, thermal energy is stored via a borehole heat exchanger in the ground.
Model predictive control (MPC) may maximize the use of BTES by achieving a dynamic interaction between the building and BTES.
However, modeling BTES for MPC is challenging, and a trade-off between model accuracy and an easy-to-solve optimal control problem (OCP) must be found.
This manuscript presents an accurate numerical model yielding an easy-to-solve linear-quadratic OCP.
\end{abstract}

\section{Introduction}
The European Union's ``Fit for 55'' initiative aims to reduce net greenhouse gas emissions by 55\% \citep{EuropeanCommission.2021}.
In 2023, the operation of heating, ventilation, and air-conditioning (HVAC) technology for buildings still accounts for approximately 30\% of global final energy consumption, making it a preferred emission reduction target \citep{InternationalEnergyAgency.2023}.

A ubiquitous CO$_2$ mitigation approach is to prevent excess thermal energy from dissipating and store it in, e.g., borehole thermal energy storage (BTES) \citep{Lee.2013}.
Thus, the storage recharges in an environmentally friendly way and, at the same time, conditions the building by providing thermal energy.
This reduces the operation and emissions of simultaneously installed fossil fuel-based HVAC technologies of the building.

The storage’s aim is a vigorous contribution to the thermal energy demand of the building and can be achieved with, e.g., model predictive control (MPC).
\citet{Drgona.2020} show that MPC may effectively increase the thermal comfort and energy savings for buildings.
MPC advantageously uses the solution of an optimal control problem (OCP), which covers model-based predictions of the building and BTES.
\citet{Verhelst.2011, Stoffel.2022}, and \citet{Kumpel.2022} thoroughly discuss MPC approaches for BTES.

Ideally, the prediction model for MPC describes the BTES system accurately and yields an easy-to-solve OCP.
Both is not always given \citep{Heim.2024}.
The presented numerical BTES system model of this manuscript closes this gap.
An affine state-space model can represent the novel model and beneficially yields an easy-to-solve quadratic program as OCP.

\subsection{Principles of BTES system operation}\label{sec:01-principles-of-btes-operation}
\begin{figure}[t]
    \centering
    \includegraphics[trim={0cm, 22cm, 13.2cm, 0cm}, clip, width=0.7\linewidth]{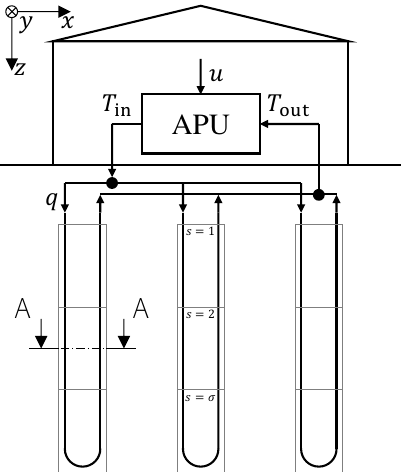}
    \caption{Schematic illustration of a BTES with $\nu = 3$ BHE connected to a building via an auxiliary power unit (APU).
    The BHE are divided into $\sigma = 3$ segments.
    Cross section A:A is given in Figure~\ref{fig:TRC-BHE-model}.
    }
    \label{fig:btes}
\end{figure}

\citet{Lee.2013} and \citet{Stober.2021} give a comprehensive introduction to BTES.
BTES typically comprises a set of borehole heat exchangers (BHE) with lengths between \qty{20}{\meter} and \qty{300}{\meter}.
It uses the BHE to exchange thermal energy with the surrounding ground (see Figure~\ref{fig:btes}), where the thermal energy is stored.
Normally, the BHE are vertically positioned in the ground and contain pipes, which guide process fluid \citep{Laferriere.2020}.
The process fluid
, a mixture of, e.g., water and glycol,
transports thermal energy from the building via an auxiliary power unit (APU) or a heat exchanger to the BHE and the ground.
The space between the pipes in the BHE is filled with backfill material, e.g., grout or bentonite.

The operation of BTES systems is based on excess thermal energy from summer and winter.
In summer, excess heat is removed from the building and stored via the process fluid and the BHE in the ground.
The BTES is charged with heat, and, consequently, the ground heats above its ambient temperature.
There, the ground's large specific heat capacity and little ambient groundwater flow keep the thermal energy until the building needs heat in winter.
Then, the stored thermal energy is taken from the ground and delivered to the building.
At the same time, the ground cools and is prepared to cool the building in summer.

\subsection{Modeling BTES systems for MPC}\label{sec-01-modeling-btes}
The prediction model for MPC is a key factor for its success.
The better the predictions, the better the MPC's decisions for the system.
Further, the model should preferably yield an easy-to-solve OCP to guarantee low solver runtimes \citep{Heim.2024}.

As the control of BTES has been an active research field for more than a decade, many models for BTES in the context of MPC have been developed.
Although all models are MPC-capable, not all models yield an easy-to-solve OCP or are reasonably accurate.
A comprehensive overview by \citet{Laferriere.2020} is briefly presented here for completeness.
Commonly, BTES system models are divided into three sub-models: a first model, which focuses on the energy transport in the BHE; a second model, which focuses on the energy transport in the ground; and a third model, which focuses on the building's thermal behavior.
The following review highlights the ground's energy transport and BHE models.

Typically, a thermal-resistance-capacitance (TRC) approach applies to the BHE model.
Corresponding resistances and capacities can be determined based on the multipole method, line-source approximations, or finite element methods.
However, many methods only focus on steady-state approximations, which neglect transient effects \citep{Laferriere.2020}.
Transient effects can be modeled using the methodology of \citet{Bauer.2011}.

Next to high-fidelity numerical models, the energy transport in the ground and the effects of the ground on the BHE can be approximated with an infinite line source and cylindrical heat source solutions.
Additionally, g-functions based on BHE's average responses of a unit-step can be used \citep{Laferriere.2020}.
These models cannot display short-term effects, such as fast-changing process fluid temperatures.
To account for that, g-functions can be adapted with, e.g., isothermal line sources.

\subsection{Contributions}\label{sec:01-contributions}
A recent study by \citet{Heim.2024} examines the use of different BTES system models for MPC.
They validate four models (a numerical model, a g-function model, a TRC model, and a hybrid model) with high-frequency measurement data from a real plant.
Based on this, they provide recommendations on which model is best suited for MPC.
The hybrid and numerical models show accurate predictions for winter and summer, whereas the numerical model is slightly better at predicting short-term effects, such as temperature changes of the process fluid.
The authors mention that MPC with their professional-software numerical model is cumbersome because the access to the model's equations is limited.

Based on the statements of \citet{Heim.2024}, we have developed a lightweight numerical model with finite volume methods, which transforms to an affine state-space model
\begin{equation}
    \label{eq:BTES-state-space}
    \xb(t_{k+1}) = \Ab \xb(t_{k}) + \Bb u(t_{k}) + \fb \, .
\end{equation}
Here, $\Ab \in \R^{n \times n}$ denotes the state matrix, $\Bb  \in \R^{n \times 1}$ the input matrix, $\fb \in \R^{n}$ the state offset vector, $\xb(t_{k}) \in \R^{n}$ the state vector, and $u(t_{k})$ the system input.
$t_{k}$ refers to the time at $t_{k} = k \Delta t \, \forall \, k\in \N $, where the discretization step size is $\Delta t$.
This model combines the highest model accuracy with an easy-to-solve OCP, which is a quadratic program.

The novel model comprises three sub-models: a numerical two-dimensional heat transport model for the ground, a set of $\nu$ vertically discretized TRC BHE models, and an APU model.
Here, the APU may be interpreted as an arbitrary interface between the BTES and the building, and it is, therefore, simplified.
The state vector~$\xb(t_{k})$ is a conjunction of the sub-models' states.

The main contributions are the development of the two-dimensional numerical ground model and the conjunction of each sub-model to one BTES system model.
Additionally, the numerical ground model is an extension of existing ground models (used for MPC), as it considers groundwater flow.
To the authors' knowledge, this is not covered by the presented models in Section~\ref{sec-01-modeling-btes}.

\section{Novel Model}\label{sec:02-novel-model}
The novel model combines three sub-models: a two-dimensional numerical ground model, a vertically discretized TRC BHE model, and a model for the APU.
The state-space representation of all sub-models,
\begin{equation}
    \label{eq:sub-models-state-space}
    \begin{split}
        \xb_{\mathrm{G}}(t_{k+1}) & = \Ab_{\mathrm{G}} \xb_{\mathrm{G}}(t_{k}) + \fb_{\mathrm{G}}(\xb_{\mathrm{B}1}(t_{k}), ..., \xb_{\mathrm{B}\nu}(t_{k})) \\
        \xb_{\mathrm{B}j}(t_{k+1}) & = \Ab_{\mathrm{B}j} \xb_{\mathrm{B}j}(t_{k}) + \fb_{\mathrm{B}j}(\xb_{\mathrm{G}}(t_{k}), \xb_{\mathrm{A}}(t_{k})) \\
        \xb_{\mathrm{A}}(t_{k+1}) & = \Ab_{\mathrm{A}} \xb_{\mathrm{A}}(t_{k}) + \Bb_{\mathrm{A}}u(t_{k}) \, ,
    \end{split}
\end{equation}
is disclosed in the following subsections.
Here, $\fb_{\mathrm{G}}(\cdot)$ and $\fb_{\mathrm{B}j}(\cdot)$ denote state offset vectors for affine systems.
The indexes $\mathrm{G}$, $\mathrm{B}$, and $\mathrm{A}$ highlight the ground, BHE, and APU model.
The state-space matrices of the BHE model are marked with the index $j$ to represent one of the $\nu$ BHE.

The BTES system model from Equation~\eqref{eq:BTES-state-space} is based on the sub-model's state-space representation from Equation~\eqref{eq:sub-models-state-space}.
The BTES system's state vector $\xb(t_{k})$ is a concatenation of $\xb_{\mathrm{A}}(t_{k})$, $\xb_{\mathrm{B}1}(t_{k})$, ..., $\xb_{\mathrm{B}\nu}(t_{k})$, and $\xb_{\mathrm{G}}(t_{k})$.
The state matrix $\Ab$ is given by the block-diagonal of the sub-model's state matrices ($\Ab_{\mathrm{A}}, \Ab_{\mathrm{B}1}, ..., \Ab_{\mathrm{B}\nu}, \Ab_{\mathrm{G}}$).
Further, note that the state offset vectors $\fb_{\mathrm{G}}(\cdot)$ and $\fb_{\mathrm{B}j}(\cdot)$ in Equation~\eqref{eq:sub-models-state-space} are a function of the states $\xb_{\mathrm{B}j}(t_{k})$, $\xb_{\mathrm{A}}(t_{k})$, and $\xb_{\mathrm{G}}(t_{k})$, which highlights the connections between the sub-models.
Consequently, the elements of the state offset vectors $\fb_{\mathrm{G}}(\cdot)$ and $\fb_{\mathrm{B}j}(\cdot)$ that depend on $\xb_{\mathrm{B}j}(t_{k})$, $\xb_{\mathrm{A}}(t_{k})$, and $\xb_{\mathrm{G}}(t_{k})$ slide partially (only the factors of the states) into the state matrix~$\Ab$ (next to the block diagonal matrices).
State offset vector entries that are not a function of the sub-models' states remain in the state offset vector~$\fb$.
The system's input matrix~$\Bb$ simply holds the input matrix~$\Bb_{\mathrm{A}}$ of the APU model.

\subsection{Ground model}\label{sec:02-Ground-model}
The ground's temperature $T(x,y,t)$ is modeled by solving the two-dimensional governing partial differential equation for heat transport in the ground \citep{Anderson.2005},
\begin{gather}
    \label{eqn:T_PDE_nonlinear}
        c_{\mathrm{g}} \frac{\partial T(x,y,t)}{\partial t} = \lambda \left( \frac{\partial^{2} T(x,y,t)}{\partial x^{2}} + \frac{\partial^{2} T(x,y,t)}{\partial y^{2}} \right) \\
        -c_{\mathrm{w}} \phi \left( v_{x}(x,y) \frac{\partial T(x,y,t)}{\partial x} + v_{y}(x,y) \frac{\partial T(x,y,t)}{\partial y} \right). \nonumber
\end{gather}
Ambient groundwater flow $v_{\{x, y\}}(\cdot)$ in $x$- and $y$-direction is considered by this equation.
$c_{g}$ and $c_{w}$ denote the volumetric heat capacity of the ground (mixture of rock and water) and groundwater.
$\lambda$ and $\phi$ represent the heat conduction coefficient and porosity of the ground.

\begin{figure}[t]
    \centering
    \includegraphics[trim={0cm, 11.4cm, 1.5cm, 0cm}, clip, width=0.7\linewidth]{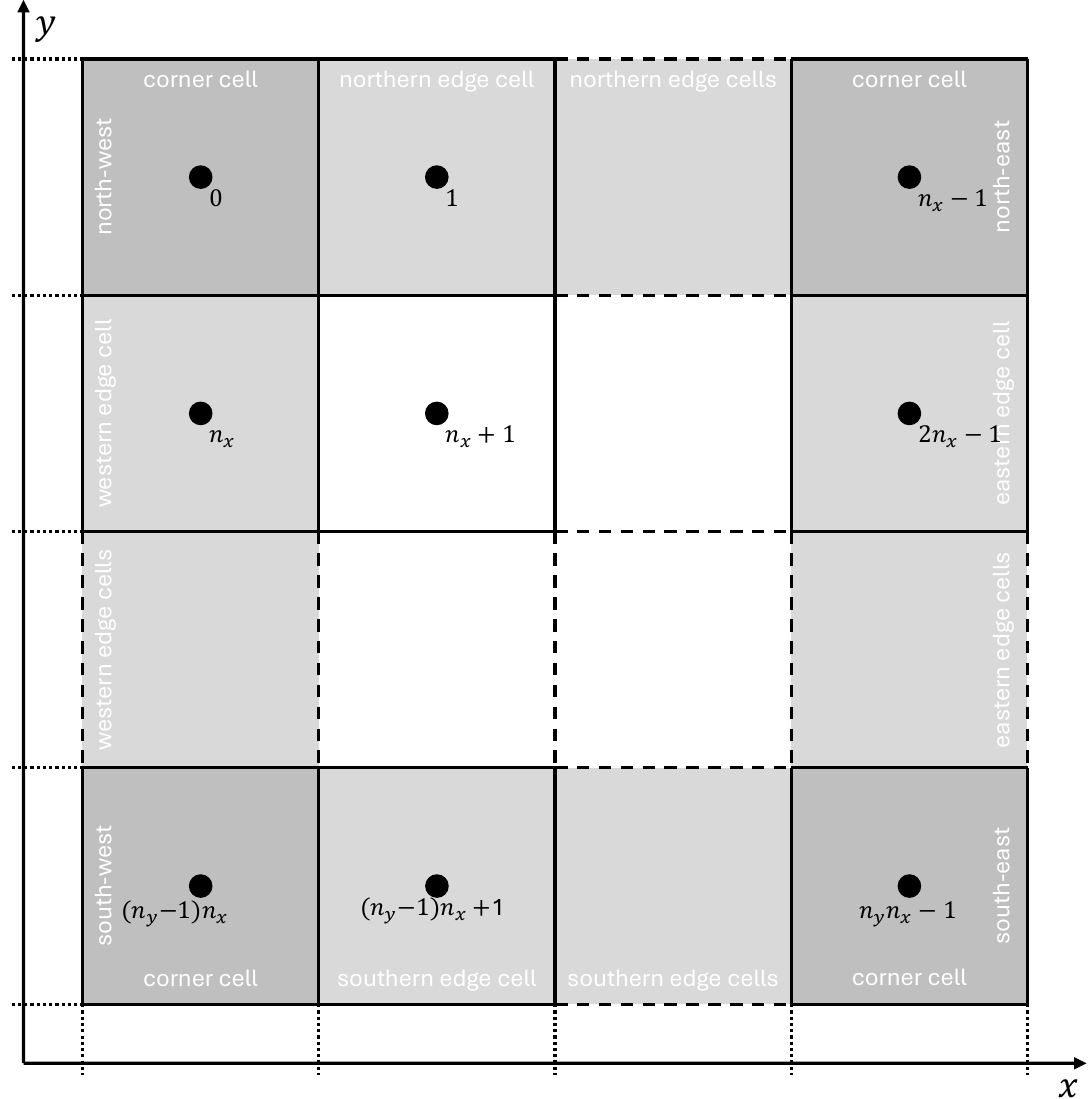}
    \caption{
        Illustration of the mesh's discretization with cell identification numbers.
        Edge and corner cells are highlighted in light and dark gray, respectively.
    }
    \label{fig:2d-mesh-number-scheme}
\end{figure}

The discretization of the spatial domain is illustrated in Figure~\ref{fig:2d-mesh-number-scheme}.
Every cell has a unique identification number, which determines the location of its center or bounds.
Edge and corner cells at the bounds of the spatial domain are highlighted in light and dark gray.
The domain is divided into $n_{x}$ and $n_{y}$ cells in $x$- and $y$-direction.

Now, for every cell, the finite volume method is applied.
The theory of the finite volume method may be studied with, e.g., \citet{Schafer.2022} or \citet{Ferziger.2002}.
Upwind differencing for convective (second addend in \eqref{eqn:T_PDE_nonlinear}) and central differencing for conductive fluxes (first addend in \eqref{eqn:T_PDE_nonlinear}) are used to achieve stability and boundedness of the numerical solution \citep{Schafer.2022}.
Implicit or explicit Euler integration applies to integrate over time.

\begin{figure}
    \centering
    \includegraphics[trim={0cm, 20cm, 9,8cm, 0.3cm}, clip, width=0.5\linewidth]{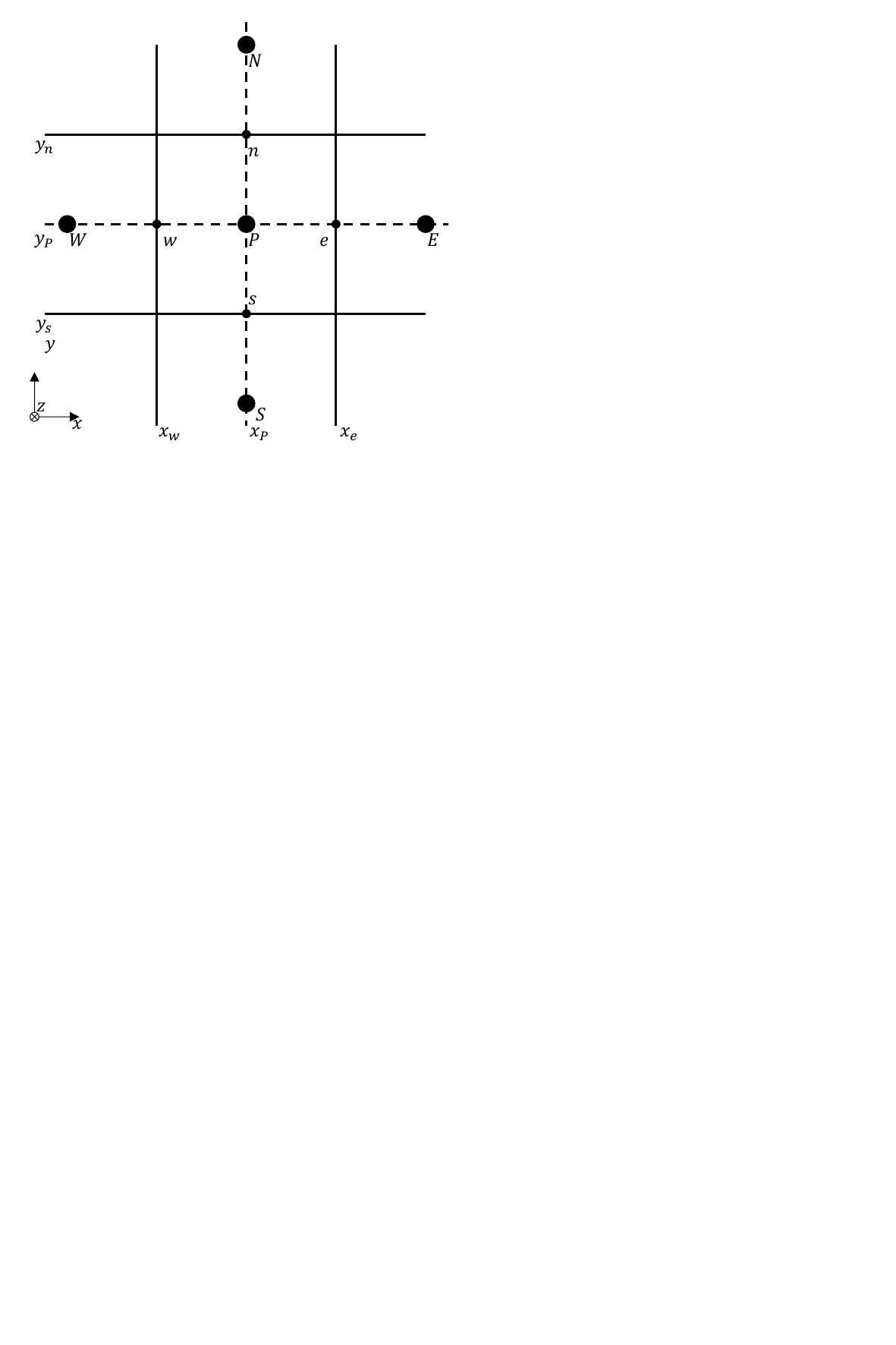}
    \caption{
        Illustration of the two-dimensional mesh in the WESN representation.
        Cell centers are marked with uppercase letters (cell center with $P$ and neighboring cell centers with north~$N$, south~$S$, west~$W$, and east~$E$).
        Its bounds are marked with lowercase letters (north~$n$, south~$s$, west~$w$, and east~$e$).
    }
    \label{fig:2d-mesh-pwne}
\end{figure}

The finite volume method for the general cell $i$ is presented in the following paragraphs.
For this, the west-east-south-north (WESN) relative cell representation from Figure~\ref{fig:2d-mesh-pwne} is used \citep{Schafer.2022, Ferziger.2002}.
Note that the cells' centers are marked with capital letters (P, W, E, S, and N) and the cell's bounds with lowercase letters (w, e, s, and n).
P refers to the center of cell $i$.

The finite volume method with explicit Euler integration yields the expression for the temperature of cell $i$
\begin{multline}
    \label{eq:fvm-T-PDE}
      c_{\mathrm{g}}
      \frac{ T_{\mathrm{P}}(t_{k+1}) -  T_{\mathrm{P}}(t_{k})}{\Delta t}
      =
      \\
    + \frac{\lambda}{\Delta x}
        \left(
            \frac{T_{\mathrm{e}}(t_{k})}{\partial x}
            - \frac{\partial T_{\mathrm{w}}(t_{k})}{\partial x}
        \right)
    + \frac{\lambda}{\Delta y}
        \left(
            \frac{T_{\mathrm{n}}(t_{k})}{\partial y}
            - \frac{\partial T_{\mathrm{s}}(t_{k})}{\partial y}
        \right)
        \\
    + \frac{c_{\mathrm{w}} \phi}{\Delta x}
        \left(
        T_{\mathrm{w}}(t_{k}) v_{x}(x_{\mathrm{w}}, y_{\mathrm{P}})
        - T_{\mathrm{e}}(t_{k}) v_{x}(x_{\mathrm{e}}, y_{\mathrm{P}})
        \right)
        \\
    + \frac{c_{\mathrm{w}} \phi}{\Delta y}
        \left(
        T_{\mathrm{s}}(t_{k}) v_{y}(x_{\mathrm{P}} , y_{\mathrm{s}})
        - T_{\mathrm{n}}(t_{k}) v_{y}(x_{\mathrm{P}}, y_{\mathrm{n}})
        \right)
    \, .
\end{multline}
Note that $T_{\{\mathrm{w, e, s, n}\}}(t_{k})$ are the temperatures at the cell's bounds (see Figure~\ref{fig:2d-mesh-pwne}).
Here, $\Delta x = x_{\mathrm{e}} - x_{\mathrm{w}}$ and $\Delta y = y_{\mathrm{n}} - y_{\mathrm{s}}$ are the edge lengths of cell $i$.

The upwind and central differencing scheme must now be applied to Equation~\eqref{eq:fvm-T-PDE}.
A general expression for this is not presented here, because it depends on the direction of the ambient groundwater flow~$v_{\{x,y\}}(\cdot)$ \citep{Schafer.2022, Ferziger.2002}.
However, after applying upwind and central differencing, Equation~\eqref{eq:fvm-T-PDE} becomes a linear function~$\mathfrak{f}$
\begin{equation}
    \label{eq:pwesn-state}
    T_{\mathrm{P}}(t_{k+1}) = \mathfrak{f}(T_{\mathrm{P}}(t_{k}), T_{\mathrm{W}}(t_{k}), T_{\mathrm{E}}(t_{k}), T_{\mathrm{S}}(t_{k}), T_{\mathrm{N}}(t_{k})) \, ,
\end{equation}
which has the center temperature~$T_{\mathrm{P}}(t_{k})$ and the temperatures of the neighboring cells W, E, S, and N as input arguments.
Note that the upwind and central differencing schemes replace the temperatures at the bounds~$T_{\{\mathrm{w, e, s, n}\}}(t_{k})$ with the temperatures at the neighboring cells' center~$T_{\{\mathrm{W, E, S, N}\}}(t_{k})$.
With this result, the affine state-space system is created.

After applying the finite volume method on all cells, boundary conditions must be added, and they manipulate Equation~\eqref{eq:pwesn-state} for the edge, corner, and cells that are located next to a BHE.
The general treatment of boundary conditions may be studied with \citet{Schafer.2022}.
A Dirichlet boundary condition imposes the ground's ambient temperature for edge and corner cells.
The energy exchange between the BHE and the ground is represented by a Neumann boundary condition, which implies the Fourier law of thermal conduction.
Accordingly, the computation of the heat flux~$Q_{\mathrm{B}j}(\xb_{\mathrm{B}j})$ between the $j$-th BHE and the ground is given by Equation~\eqref{eq:BHE-Ground-heat-flow} in the following section because the BHE model should be introduced first.

The conjunction of \eqref{eq:pwesn-state} for all cells transforms to an affine state-space system (as already presented in \eqref{eq:sub-models-state-space})
\begin{equation*}
    \xb_{\mathrm{G}}(t_{k+1}) = \Ab_{\mathrm{G}} \xb_{\mathrm{G}}(t_{k}) + \fb_{\mathrm{G}}(\xb_{\mathrm{B}1}(t_{k}), ..., \xb_{\mathrm{B}\nu}(t_{k})) \, ,
\end{equation*}
where $\xb_{\mathrm{G}}(t_{k}) = [T_{0}^{k}, ..., T_{n_{x}n_{y}-1}^{k}] \in \R^{n_{x}n_{y}}$.
$T_{i}^{k}$ denotes the temperature at the center of the cell~$i$ at time~$t_{k}$.

\subsection{Borehole heat exchanger model}\label{sec:02-borehole-heat-exchanger}
\begin{figure}[t]
    \centering
    \includegraphics[trim={0cm, 23.3cm, 13.8cm, 0cm}, clip, width=0.5\linewidth]{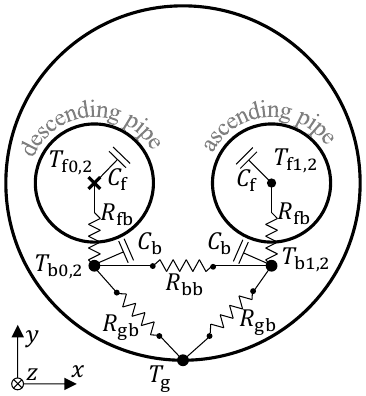}
    \caption{Cross section A:A from Figure~\ref{fig:btes} shows the delta-circuit TRC BHE model (adapted from \citep{Laferriere.2020}).
    }
    \label{fig:TRC-BHE-model}
\end{figure}

The vertically discretized TRC model for a single U-tube BHE is based on \citet{Laferriere.2020} and \citet{Bauer.2011}.
In the vertical direction, the BHE is divided into $\sigma$ segments.
For each segment~$s$, the model approximates the energy transport with a delta-circuit network of thermal resistances and capacities (see Figure~\ref{fig:TRC-BHE-model}).
The resistances and capacities are determined by the methodology of \citet{Bauer.2011}.
The model covers for each segment~$s \in \{1, ..., \sigma \}$ the process fluid temperature~$T_{\mathrm{f\{0,1\}},s}(t_{k})$ in the descending (with index~$0$) and ascending (with index~$1$) pipe, the backfill temperatures~$T_{\mathrm{b\{0,1\}},s}(t_{k})$ close to the descending and ascending pipes, and the ground temperature at the borehole wall~$T_{\mathrm{g}}(t_{k})$.
In the descending pipe, the process fluid flows in positive $z$-direction; in the ascending pipe, vice versa.

The process fluid temperatures~$T_{\mathrm{f\{0,1\}},s}(t_{k})$ depend on the enthalpy transport due to the process fluid flow~$q$ and heat conduction with the backfill material, which is physically described with the resistance~$R_{\mathrm{fb}}$.
Thus, the process fluid temperature for the descending pipe is computed with explicit Euler by
\begin{multline}
    C_{\mathrm{w}} \frac{T_{\mathrm{f0},s}(t_{k+1}) - T_{\mathrm{f0},s}(t_{k})}{\Delta t} = \frac{T_{\mathrm{b0},s}(t_{k}) - T_{\mathrm{f0},s}(t_{k})}{R_{\mathrm{fb}}} \\
    + \frac{q c_{\mathrm{w}}}{l} \left(T_{\mathrm{f0},s-1}(t_{k}) - T_{\mathrm{f0},s}(t_{k}) \right) \, ,
\end{multline}
where $l$ denotes the segment length, $R_{\mathrm{fb}}$ the thermal resistance between the fluid and the backfill material of the BHE, $C_{\mathrm{w}}$ is the thermal capacity per unit length, and $T_{\mathrm{f0},s-1}(t_{k})$ is the temperature of the segment above (see Figure~\ref{fig:btes}).
An equation for the ascending pipe is omitted for brevity (see \citet{Laferriere.2020}).

The backfill temperatures~$T_{\mathrm{b\{0,1\}},s}(t_{k})$ depend on heat conduction with the process fluid, the borehole wall temperature~$T_{\mathrm{g}}(t_{k})$, and each other.
This is mathematically represented by
\begin{multline}
    C_{\mathrm{b}} \frac{T_{\mathrm{b0},s}(t_{k+1}) - T_{\mathrm{b0},s}(t_{k})}{\Delta t} = \frac{T_{\mathrm{f0},s}(t_{k}) - T_{\mathrm{b0},s}(t_{k})}{R_{\mathrm{fb}}} \\
    + \frac{T_{\mathrm{b1},s}(t_{k}) - T_{\mathrm{b0},s}(t_{k})}{R_{\mathrm{bb}}} + \frac{T_{\mathrm{g}}(t_{k}) - T_{\mathrm{b0},s}(t_{k})}{R_{\mathrm{gb}}}
\end{multline}
for the backfill temperature~$T_{\mathrm{b}0,s}(t_{k})$ close to the descending pipe.
Here, $C_{\mathrm{b}}$ denotes the backfill materials specific heat capacity per unit length, $R_{\mathrm{bb}}$ the thermal resistances between backfill close to the descending and ascending pipes, and $R_{\mathrm{gb}}$ the thermal resistance between backfill and the borehole wall.
An equation for the ascending pipe is omitted for brevity (see \citet{Laferriere.2020}).

The borehole wall temperature~$T_{\mathrm{g}}(t_{k})$ represents the average ground temperature at the center of all neighboring BHE cells,
\begin{equation}
    T_{\mathrm{g}}(t_{k}) = \frac{1}{4} \left( T_{\mathrm{E}}(t_{k}) + T_{\mathrm{W}}(t_{k}) + T_{\mathrm{S}}(t_{k}) + T_{\mathrm{N}}(t_{k}) \right) \, .
\end{equation}

Presented equations hold for all BHE and segments~$s$ with a few exceptions.
The enthalpy flux into the descending pipe of the top-level segment (i.e., $s=0$ refer to Figure~\ref{fig:btes}) of the BHE is determined by the BTES' inlet temperature~$T_{\mathrm{in}}(t_{k})$.
This temperature is given by the state of the APU $\xb_{\mathrm{A}}(t_{k})$ and appears in the state offset vector~$\fb_{\mathrm{B}j}(\cdot)$.
Please refer to Section~\ref{sec:02-auxiliary-power-unit} for further information.
The enthalpy flux into the ascending pipe of the lowest segment (i.e., $s=\sigma$ refer to Figure~\ref{fig:btes}) of the BHE is determined by the process fluid's temperature of the descending pipe in the same segment.

Now, all equations of this subsection are brought to an affine state-space model (see \eqref{eq:sub-models-state-space})
\begin{equation*}
    \xb_{\mathrm{B}j}(t_{k+1}) = \Ab_{\mathrm{B}j} \xb_{\mathrm{B}j}(t_{k}) + \fb_{\mathrm{B}j}(\xb_{\mathrm{G}}(t_{k}), \xb_{\mathrm{A}}(t_{k}))
\end{equation*}
for a better handling of the MPC algorithm, where $\xb_{\mathrm{B}j}(t_{k})$ holds the temperatures $T_{\mathrm{f\{0,1\}},s}(t_{k})$ and $T_{\mathrm{b\{0,1\}},s}(t_{k})$ for all segments~$s$.

The heat transfer rate between the $j$-th BHE and the ground is given by
\begin{equation}
    \label{eq:BHE-Ground-heat-flow}
    Q_{\mathrm{B}j}(t_{k+1}) = \frac{1}{R_{\mathrm{gb}}} \left(\frac{1}{\sigma} \sum_{s=1}^{\sigma} \sum_{r=0}^{1} T_{\mathrm{b}r,s}(t_{k})
    - 2 T_{\mathrm{g}}(t_{k}) \right).
\end{equation}
which is a linear relation between the borehole wall temperature~$T_{\mathrm{g}}(t_{k})$, the backfill temperatures of all segments, and the thermal resistance~$R_{\mathrm{gb}}$.

\subsection{Auxiliary power unit model}\label{sec:02-auxiliary-power-unit}
The APU thermally connects the BTES with the building.
To keep it simple, it is assumed that the APU can add and remove thermal energy~$u(t_{k})$ from the process fluid.
$u(t_{k})$ is the system input for the complete BTES system model.
The physical relation between the BTES' inlet~$T_{\mathrm{in}}(t_{k})$ and outlet temperature~$T_{\mathrm{out}}(t_{k})$ is
\begin{equation}
    \label{eq:apu}
    T_{\mathrm{in}}(t_{k+1}) = T_{\mathrm{out}}(t_{k}) + \frac{1}{\nu q c_{\mathrm{w}}} u(t_{k}) \, .
\end{equation}
This equation transforms to a linear state-space system according to Equation~\eqref{eq:sub-models-state-space}.
The state matrix~$\Ab_{\mathrm{A}} \in \R^{2\times2}$ and input matrix~$\Bb_{\mathrm{A}} \in \R^{2 \times 1}$.

\section{Numerical studies}\label{sec:03-numerical-studies}
This section is divided into three subsections.
The first tests the TRC BHE model against real data published by \citet{Beier.2011}.
The second presents simulation results of the BTES system model with nine BHE, and the third demonstrates the closed-loop MPC capabilities of the novel model.

The used parameters for the numerical studies are disclosed in the following paragraphs.
Unless it is remarked, these parameters are valid for all subsections.
Note that all parameters are based on values from the literature to prove the functionality of the model.
However, they are not linked to a real site.

By assumption, groundwater flows five millimeters per hour, $v_{\{x, y\}} = \qty{1.3889}{\micro\meter\per\second}$, and the ambient ground temperature is \qty{295.15}{\kelvin} (\qty{22}{\degreeCelsius}).
The ground has a porosity of $\phi = 0.8$ and a heat conduction coefficient of $\lambda = \qty{2.3}{\watt\per\meter\kelvin}$.
The specific heat capacity for the ground~$c_{\mathrm{g}}$ and groundwater~$c_{\mathrm{w}}$ are
\qtylist{2300000; 4200000}{\joule\per\cubic\meter\kelvin}.

The spatial domain of the ground model is discretized with a non-uniform structured mesh that imposes a fine meshing close to the nine BHE for better accuracy.
Here, the domain's size is $\qty{20}{\meter}$ in $x$- and $y$-direction.
The fine/coarse cells have an edge length of \num[exponent-mode=input]{0.2}/\qty{1}{\meter}, which yields $n_{x} = 47$ and $n_{y} = 47$ cells in the $x$- and $y$-direction.
In total, \num[exponent-mode=input]{2209} cells represent the spatial domain.
The discretization step size is $\Delta t = \qty{15}{\second}$.

The process fluid flows with $q = \qty{0.1974}{\kilogram\per\second}$ through the $\sigma = 3$ segments with a segment length~$l = \qty{3.66}{\meter}$.
The specific heat capacity per unit length for the process fluid and the backfill material are $C_{\mathrm{w}} = \qty{2452.7037}{\joule\per\meter\kelvin}$ and $C_{\mathrm{b}} = \qty{20361.6610}{\joule\per\meter\kelvin}$.
Thermal resistances are computed using the methodology of \citet{Bauer.2011}, which yields \qty{0.261}{\watt\per\meter\kelvin}, \qty{0.4538652673363449}{\watt\per\meter\kelvin}, and \qty{0.06931151010684597}{\watt\per\meter\kelvin} for $R_{\mathrm{fb}}$, $R_{\mathrm{bb}}$, and $R_{\mathrm{gb}}$.

The complete BTES system model has $n = \num[exponent-mode=input,round-mode=none]{2319}$ states.
Nine BHE are placed in the spatial domain with a distance of \qty{2}{\meter} in the $x$- and $y$-direction (refer to Figure~\ref{fig:03-heat-map}).

\subsection{BHE model test}\label{sec:03-BHE-model-test}
The BHE model approximates the reference dataset for a single U-tube BHE from \citet{Beier.2011}.
The data was recorded during a thermal response test; ground temperatures are not recorded.
This section assumes no groundwater flow and one BHE in the ground.

\begin{figure}
    \centering
    \includegraphics[trim={0.6cm, 6cm, 0.8cm, 0.5cm}, clip, width=0.7\linewidth]{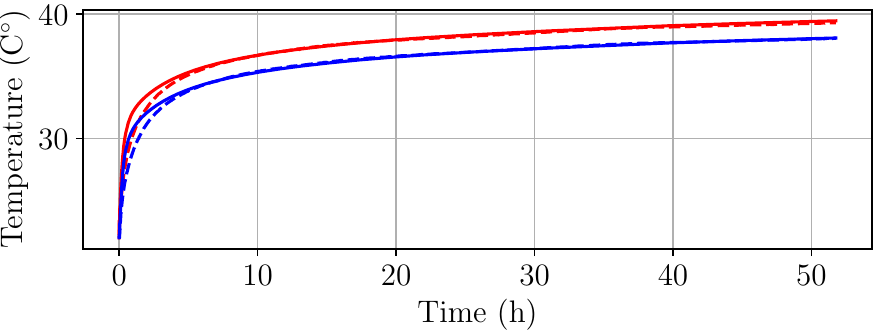}
    \caption{Result of the BHE simulation (solid) compared to real data (dashed) from \citet{Beier.2011}.
        The red lines represent BTES' inlet temperature~$T_{\mathrm{in}}(t_{k})$, and the blue line represents BTES' outlet temperature~$T_{\mathrm{out}}(t_{k})$.
    }
    \label{fig:Beier-BHE}
\end{figure}

Simulated (solid) and measured (dashed) BTES' inlet~$T_{\mathrm{in}}(t_{k})$ (red) and outlet (blue) temperature~$T_{\mathrm{out}}(t_{k})$ over time are given by Figure~\ref{fig:Beier-BHE}.
It is noticeable that the temperature of the process fluid rises during the experiment from \qty{295.15}{\kelvin} (\qty{22}{\degreeCelsius}) to about \qty{313.15}{\kelvin} (\qty{40}{\degreeCelsius}).
The inlet temperature~$T_{\mathrm{in}}(t_{k})$ remains higher than the outlet temperature~$T_{\mathrm{out}}(t_{k})$, which shows that thermal energy is added to the ground; the BTES is charged with heat.
For the \qty{52}{\hour}, the mean error between simulation and real data is \qty{0.182870}{\kelvin} (with standard deviation \qty{0.40319}{\kelvin}) for the inlet temperature~$T_{\mathrm{in}}(t_{k})$ and \qty{0.08255}{\kelvin} (with a standard deviation of \qty{0.41135}{\kelvin}) for the outlet temperature~$T_{\mathrm{out}}(t_{k})$.

\subsection{BTES simulation}\label{sec:03-btes-simulation}
For this simulation, the APU heats the process fluid with \qty{4500}{\watt} (\qty{500}{\watt} for each BHE) for \qty{26}{\hour}.
The ground is at ambient temperature at the beginning of the simulation.

\begin{figure}
    \centering
    \includegraphics[width=0.7\linewidth]{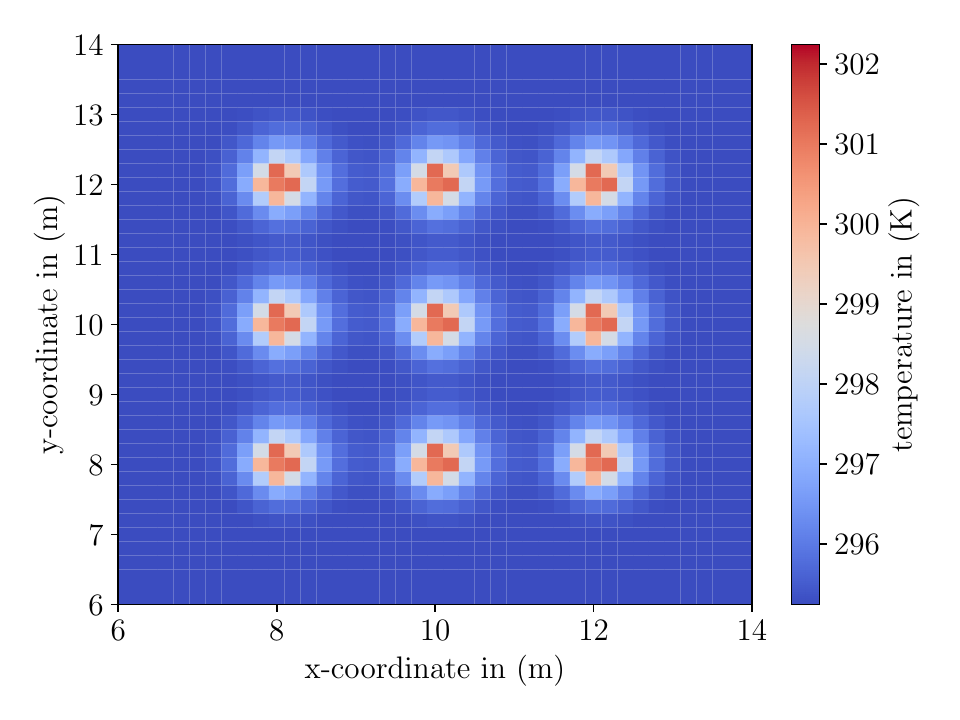}
    \caption{Heat map of BTES simulation after \qty{26}{\hour}.}
    \label{fig:03-heat-map}
\end{figure}

Figure~\ref{fig:03-heat-map} illustrates the ground's temperature at the end of the simulation.
The ground close to the BHE heated up to about \qty{302}{\kelvin}.
Due to the ambient groundwater flow (going from southwest to northeast), the ground's temperature northeast of the BHE is hotter than southwest.

\subsection{Model predictive control}
This section briefly demonstrates that an OCP for a simple tracking MPC algorithm with the novel BTES system model is easy-to-solve and requires only a little computational power.
The MPC algorithm tracks the building's energy demand for 24 hours with a prediction horizon~$H = 80$, which is equivalent to \qty{20}{\minute}.
Basics about (tracking) MPC may be studied with, e.g., \citet{Rawlings.2022}.

As time proceeds, the tracking MPC computes an optimal input sequence by recursively solving the OCP every $t_{k}$.
The input sequence~$\hat{\ub}_{H}$ holds the predicted inputs~$\hat{u}(\tau_{\kappa}) \, \forall \kappa \in \{0, ..., H - 1 \}$, where $\tau_{\kappa} = t_{k} + \kappa \Delta t$.
Only the first element~$\hat{u}(\tau_{0})$ of the input sequence~$\hat{\ub}_{H}$ is applied onto the system ($\hat{u}(\tau_{0}) = u(t_{k})$).
The OCP,
\begin{align}
    & \min_{\hat{\ub}_{H}} \sum_{\kappa = 0}^{H - 1}  \|\uh(\tau_{\kappa}) - \yh_{\mathrm{ref}}(\tau_{\kappa}) \|_{R}^{2} + \| \uh(\tau_{\kappa}) - \uh(\tau_{\kappa -1}) \|_{Q}^{2} \label{eq:mpc} \\
    & \text{subject to} \nonumber
\end{align}
\begin{align*}
    \xbh(\tau_{0}) & = \xb(t_{k}) \\
    \xbh(\tau_{\kappa+1}) & = \Ab \xbh(\tau_{k}) + \Bb \uh(\tau_{k}) + \fb \quad \forall \kappa \in \{0, ..., H-1\} \\
    \xbh(\tau_{\kappa}) & \in \mathbb{X} \quad \forall \kappa \in \{0, ..., H\} \\
    \uh(\tau_{\kappa}) & \in \mathbb{U} \quad \forall \kappa \in \{0, ..., H-1\} \, ,
\end{align*}
is an easy-to-solve quadratic program and is initialized by the BTES system's state~$\xb(t_{k})$.
The controller tracks a predicted energy demand sequence~$\hat{y}_{\mathrm{ref}}(\tau_{\kappa})$ with constant heat demand for $\qty{5}{\minute}$ between \qty{-500}{\watt} and \qty{-1000}{\watt} (representing winter).
The tracking cost is weighted with the parameter $R = 0.1$, and the predicted power output of BTES system is given by $\uh(\tau_{k})$.
By assumption and for brevity, disturbances on the building's energy demand are neglected.
The second cost addend of the cost function is used to avoid a volatile input sequence and is weighted with $Q = 0.01$.
Note that $\hat{u}(\tau_{-1}) = u(t_{k-1})$.
$\mathbb{X}$ and $\mathbb{U}$ are convex sets, which constrain the predicted state~$\hat{\xb}(t_{k})$ and input~$\hat{u}(t_{k})$, respectively.
For the numerical study, the state constraints allow predicted states~$\hat{\xb}(t_{k})$ between \qty{273.15}{\kelvin} ($\qty{0}{\degreeCelsius}$) and \qty{303.15}{\kelvin} ($\qty{30}{\degreeCelsius}$).
APU's input~$u(t_{k})$ is limited to $\qty{ \pm 1000}{\watt}$.

\begin{figure}
    \centering
    \includegraphics[trim={0.6cm, 6cm, 0.8cm, 0.5cm}, clip, width=0.7\linewidth]{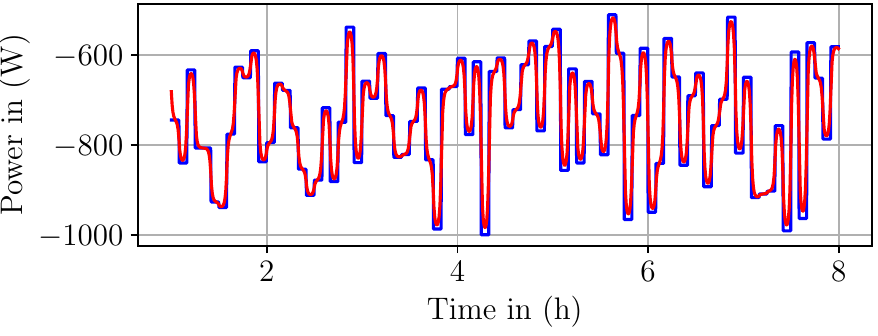}
    \caption{Delivered power~$y$ by the BTES system (red) compared to building's energy demand~$y_{\mathrm{ref}}$ (blue) over time.
    This figure represents only \qty{7}{\hour} for clarity.
    }
    \label{fig:03-mpc}
\end{figure}

The real power output~$y$ by the BTES system (red) is compared to the real building's heat demand~$y_{\mathrm{ref}}$ (blue) for \qty{7}{\hour} in Figure~\ref{fig:03-mpc}.
The BTES system can follow the heat demand of the building.
Every \qty{5}{\minute}, when the energy demand changes, the BTES system prepares itself and starts a transition phase from the current to the new energy demand.

Gurobi's average runtime for the OCP is \qty[exponent-mode=input]{0.215145}{\second} on a Debian 12 machine with an Intel Core i5-4690 and 16 GB RAM \citep{GurobiOptimization.2023}.
In the context of the discretization step size~$\Delta t$ and the prediction horizon~$H$, the OCP is easy-to-solve and requires low runtimes.

\section{Conclusion}\label{sec:04-conclusion-and-outlook}
BTES may become an important player in reducing HVAC technology's greenhouse gas emissions.
MPC is well-known and a promising control algorithm for this application.
However, MPC relies on profound prediction models for BTES systems.
This paper presents a novel BTES system model, combining model accuracy using a numerical modeling approach and an easy-to-solve OCP, as the model can be transformed to an affine state-space model.
The model design is given in Section~\ref{sec:02-novel-model}, and three numerical studies are presented in Section~\ref{sec:03-numerical-studies}.

\bibliography{literature}

\begin{thebibliography}{17}
\providecommand{\natexlab}[1]{#1}
\providecommand{\url}[1]{\texttt{#1}}
\expandafter\ifx\csname urlstyle\endcsname\relax
  \providecommand{\doi}[1]{doi: #1}\else
  \providecommand{\doi}{doi: \begingroup \urlstyle{rm}\Url}\fi

\bibitem[Anderson(2005)]{Anderson.2005}
Mary~P. Anderson.
\newblock {Heat as a ground water tracer}.
\newblock \emph{Ground Water}, 43\penalty0 (6), 2005.
\newblock ISSN 0017-467X.

\bibitem[Bauer et~al.(2011)Bauer, Heidemann, M{\"u}ller-Steinhagen, and Diersch]{Bauer.2011}
D.~Bauer, W.~Heidemann, H.~M{\"u}ller-Steinhagen, and H.-J.~G. Diersch.
\newblock {Thermal resistance and capacity models for borehole heat exchangers}.
\newblock \emph{Int. J. Energ. Res.}, 35\penalty0 (4):\penalty0 312--320, 2011.

\bibitem[Beier et~al.(2011)Beier, Smith, and Spitler]{Beier.2011}
Richard~A. Beier, Marvin~D. Smith, and Jeffrey~D. Spitler.
\newblock {Reference data sets for vertical borehole ground heat exchanger model and thermal response test analysis}.
\newblock \emph{Geothermics}, 40:\penalty0 79--85, 2011.
\newblock ISSN 03756505.

\bibitem[Drgo{\v{n}}a et~al.(2020)Drgo{\v{n}}a, Arroyo, {Cupeiro Figueroa}, Blum, Arendt, Kim, Oll{\'e}, Oravec, Wetter, Vrabie, and Helsen]{Drgona.2020}
J{\'a}n Drgo{\v{n}}a, Javier Arroyo, Iago {Cupeiro Figueroa}, David Blum, Krzysztof Arendt, Donghun Kim, Enric~Perarnau Oll{\'e}, Juraj Oravec, Michael Wetter, Draguna~L. Vrabie, and Lieve Helsen.
\newblock {All you need to know about model predictive control for buildings}.
\newblock \emph{Annu. Rev. Control}, 50, 2020.
\newblock ISSN 13675788.

\bibitem[{European Commission}(2021)]{EuropeanCommission.2021}
{European Commission}.
\newblock {Fit for 55: delivering the EU's 2030 Climate Target on the way to climate neutrality}, 2021.

\bibitem[Ferziger and Peri{\'c}(2002)]{Ferziger.2002}
Joel~H. Ferziger and Milovan Peri{\'c}.
\newblock \emph{{Computational Methods for Fluid Dynamics}}.
\newblock Springer, Berlin, DEU, 3rd edition, 2002.
\newblock ISBN 978-3-540-42074-3.

\bibitem[{Gurobi Optimization L.L.C.}(2025)]{GurobiOptimization.2023}
{Gurobi Optimization L.L.C.}
\newblock {Gurobi optimizer reference manual}.
\newblock https://www.gurobi.com, 2025.
\newblock Accessed: 10. Apr. 2025.

\bibitem[Heim et~al.(2024)Heim, Stoffel, D{\"u}ber, Knapp, K{\"u}mpel, M{\"u}ller, and Klitzsch]{Heim.2024}
Elisa Heim, Phillip Stoffel, Stephan D{\"u}ber, Dominique Knapp, Alexander K{\"u}mpel, Dirk M{\"u}ller, and Norbert Klitzsch.
\newblock {Comparison of simulation tools for optimizing borehole heat exchanger field operation}.
\newblock \emph{Geotherm. Energy}, 12\penalty0 (1), 2024.

\bibitem[{International Energy Agency}(2023)]{InternationalEnergyAgency.2023}
{International Energy Agency}.
\newblock {Tracking clean energy progress 2023: Assessing critical energy technologies for global clean energy transitions}.
\newblock https://www.iea.org/reports/tracking-clean-energy-progress-2023, 2023.
\newblock (Accessed: 10. Apr. 2025).

\bibitem[K{\"u}mpel et~al.(2022)K{\"u}mpel, Stoffel, and M{\"u}ller]{Kumpel.2022}
Alexander K{\"u}mpel, Phillip Stoffel, and Dirk M{\"u}ller.
\newblock {Development of a long-term operational optimization model for a building energy system supplied by a gerothermal field}.
\newblock \emph{J. Therm. Sci.}, 31\penalty0 (5), 2022.

\bibitem[Laferri{\`e}re et~al.(2020)Laferri{\`e}re, Cimmino, Picard, and Helsen]{Laferriere.2020}
Alex Laferri{\`e}re, Massimo Cimmino, Damien Picard, and Lieve Helsen.
\newblock {Development and validation of a full-time-scale semi-analytical model for the short- and long-term simulation of vertical geothermal bore fields}.
\newblock \emph{Geothermics}, 86, 2020.
\newblock ISSN 03756505.

\bibitem[Lee(2013)]{Lee.2013}
Kun~Sang Lee.
\newblock \emph{{Underground thermal energy storage: Green energy and technology}}.
\newblock Springer, London, GBR, 2013.

\bibitem[Rawlings et~al.(2022)Rawlings, Mayne, and Diehl]{Rawlings.2022}
James~B. Rawlings, David~Q. Mayne, and Moritz~M. Diehl.
\newblock \emph{{Model predictive control: Theory, computation, and design}}.
\newblock {Nob Hill Publishing}, Santa Barbara, CA, USA, 2022.

\bibitem[Sch{\"a}fer(2022)]{Schafer.2022}
M.~Sch{\"a}fer.
\newblock \emph{{Computational engineering: Introduction to numerical methods}}.
\newblock Springer, Cham, CHE, 2022.

\bibitem[Stober and Bucher(2021)]{Stober.2021}
Ingrid Stober and Kurt Bucher.
\newblock \emph{{Geothermal energy}}.
\newblock Springer, Cham, CHE, 2021.

\bibitem[Stoffel et~al.(2022)Stoffel, K{\"u}mpel, and M{\"u}ller]{Stoffel.2022}
Phillip Stoffel, Alexander K{\"u}mpel, and Dirk M{\"u}ller.
\newblock {Cloud-based optimal control of individual borehole heat exchangers in a geothermal field}.
\newblock \emph{J. Therm. Sci.}, 31\penalty0 (5), 2022.

\bibitem[Verhelst and Helsen(2011)]{Verhelst.2011}
Clara Verhelst and Lieve Helsen.
\newblock {Low-order state space models for borehole heat exchangers}.
\newblock \emph{HVAC{\&}R Res.}, 17\penalty0 (6), 2011.
\newblock ISSN 1078-9669.

\end{thebibliography}

\end{document}